\documentclass[11pt,paper]{JHEP}

\usepackage{amssymb}
\usepackage{amsmath}
\usepackage{amsfonts}

\allowdisplaybreaks[1]
\def\half{\tfrac{1}{2}}

\def\Tr{\,{\rm Tr}\,}
\def\makeatletter{\catcode`\@=11}
\makeatletter
\def\mathbox#1{\hbox{$\m@th#1$}}%
\def\math@ccstyles#1#2#3#4#5#6#7{{\leavevmode
      \setbox0\mathbox{#6#7}%
      \setbox2\mathbox{#4#5}%
      \dimen@ #3%
      \baselineskip\z@\lineskiplimit#1\lineskip\z@
      \vbox{\ialign{##\crcr
             \hfil \kern #2\box2 \hfil\crcr
             \noalign{\kern\dimen@}%
             \hfil\box0\hfil\crcr}}}}
\def\mathaccstyles{\math@ccstyles\maxdimen}
\def\maththroughstyles{\math@ccstyles{-\maxdimen}}
\def\unitmatrixDT%
 {\maththroughstyles{.45\ht0}\z@\displaystyle {\mathchar"006C}\displaystyle 1}
\def\leftrightarrowfill{$\mathsurround0pt \mathord\leftarrow
       \mkern-6mu\cleaders\hbox{$\mkern-2mu \mathord- \mkern-2mu$}\hfill
       \mkern-6mu \mathord\rightarrow$}

\def\overleftrightarrow#1{\vbox{\ialign{##\crcr
        \leftrightarrowfill\crcr\noalign{\kern-1pt\nointerlineskip}%
        $\hfil\displaystyle{#1}\hfil$\crcr}}}

\def\unitm#1{\unitmatrixDT_{#1}}

\newcommand{\w}{\wedge}

\newcommand{\dd}{{\rm d}}
\newcommand{\ve}{\varepsilon}
\newcommand{\E}{{\rm e}}

\numberwithin{table}{section}

\title{Group Manifold Reduction of Dual $N=1$ $d=10$ Supergravity}
\author{Mees de Roo, Martijn G.C.~Eenink, Dennis B.~Westra and\\
     Centre for Theoretical Physics\\
     Nijenborgh 4, 9747 AG Groningen,
     The Netherlands\\
     E-mail: \email{m.de.roo@rug.nl,
     m.g.c.eenink@rug.nl,\\ d.b.westra@rug.nl}}
\author{Sudhakar Panda\\
     Harish-Chandra Research Institute \\
     Chatnag Road, Jhusi, Allahabad 211019, India\\
     E-mail: \email{panda@mri.ernet.in}}

\preprint{\hepth{0503059}\\UG-05/02}

\abstract{We perform a group manifold reduction of the dual
version of $N=1$ $d=10$ supergravity to four dimensions. The
effects of the 3- and 4-form gauge fields in the resulting
gauged $N=4$ $d=4$ supergravity are studied in particular. The example of
the group manifold $SU(2)\times SU(2)$ is worked out in detail, and
we compare for this case the four-dimensional scalar potential
with gauged $N=4$ supergravity.
}

\keywords{group manifold reduction, supergravity}   

\begin{document}

\section{Introduction\label{Intro}}

Four-dimensional gauged supergravity theories are candidates
for a fundamental description of cosmological inflation. The gauging generates
a potential for the scalar fields, which can be compared to the
requirements for inflation or for accelerated expansion in general.
In this context we have been interested in particular in $N=4$ supergravity
\cite{DWP,DWPT}. One reason is that gauged $N=4$ supergravity can indeed
generate a potential with an extremum that corresponds to a positive
cosmological constant, as was first shown in \cite{GZ}. 
However, a recent analysis showed that the
potential of \cite{GZ} does not satisfy slow-roll conditions \cite{KLPS}. This
example is only one of a large class of $N=4$ models which can be
obtained in $N=4$ supergravity by introducing suitable $SU(1,1)$ angles
\cite{MdRPW1}. In \cite{DWPT} all situations where the gauged supergravity
is coupled to six additional Yang-Mills multiplets, with all possible (compact
and noncompact) semi-simple gauge groups, were analyzed. Although we
found examples with a positive extremum of the scalar potential, these
cases were not stable against fluctuations of all scalar fields.

The field content studied in \cite{DWPT} can be obtained by dimensional
reduction of $N=1$ $d=10$ supergravity, which is a second reason
for our interest in $N=4$ supergravity.  Reduction over a torus
corresponds to the ungauged supergravity theory. Scherk-Schwarz
reductions \cite{SS} or flux reductions \cite{Polchinski} will
produce a supergravity in four dimensions in which
a non-semisimple group is gauged. However, not all
four-dimensional cases can be obtained in this way, since it is not
known how to embed the dependence on the $SU(1,1)$ angles in the
reduction procedure. Further versions of $N=4$ supergravity were
obtained in \cite{TZ} by gauging directly in four dimensions, and
in \cite{AFV,AFLV} by reduction of IIB supergravity over an orbifold.
The relation between these different versions requires further
elucidation.

As a first step towards a more complete understanding of gauged $N=4$
supergravity we consider in this paper the  reduction
of the dual form \cite{Cham} of $N=1$ $d=10$ supergravity over a
compact group manifold. The consistency of this procedure is
guaranteed, see \cite{CGLP,pons} for an extensive
discussion and references on this topic.
We will limit ourselves to the bosonic sector of the theory.
The dual version contains a 6-form gauge field as the dual of the 2-form of the
standard $N=1$ theory. The 6-form gauge field reduces to 0-,1-,2-,3- and
4-form gauge fields in four dimensions. In this paper we are particularly
interested in the effect of the 3- and 4-forms. In previous
reductions of this dual version \cite{Cham,Bin,SSen}, see also \cite{Sen}
for a review, the role of these forms was not considered in all generality.
Our reduction will follow the lines of \cite{KM}, where a similar
but more elaborate analysis of the 2-form reduction was done.

This paper is organized as follows. In Section \ref{sugra} we perform the
reduction of the dual $N=1$ supergravity without additional vector
multiplets. The result of this reduction is sufficient for an analysis
of the degrees of freedom which arise from the 6-form gauge field. In Section
\ref{analysis1} we will present a cohomology technique to obtain the
massless and massive degrees of freedom for a compact group manifold.
In Section \ref{YM} we extend the results of Section \ref{sugra} by adding 
additional Yang-Mills fields in ten dimensions.
We then analyze the scalar potential in four dimensions
of the complete matter coupled system in Section \ref{potential}.
In Section \ref{conclusions} we will compare 
with gauged $N=4$ supergravity in four dimensions.

\section{Dual $N=1$ $D=10$ supergravity and Scherk-Schwarz reduction\label{sugra}}

The bosonic fields of $N=1$ supergravity in ten
dimensions are the metric $G$, the dilaton $\Phi$ and the
2-form potential $B^{(2)}$. These can be coupled to Yang-Mills
fields $A^I$, with gauge group $\mathcal{G}_{\rm YM}$. The action in the string
frame takes on the form:
\begin{equation}
  S_{\textrm{2-form}} 
  = \int \E^{-\Phi}
  \Big(
    R\,{\star\mathbf{1}} + {\star\dd\Phi}\wedge \dd\Phi
    -\tfrac{1}{2}{\star H^{(3)}}\wedge H^{(3)}
    -\Tr\,{\star F(A)}\wedge F(A)
  \Big)\,.
\label{2-form-action}
\end{equation}
The gauge invariant three-form field strength reads
\begin{equation}
  H^{(3)}= \dd B^{(2)}-\Tr \big( A\wedge\dd A + \tfrac{2}{3}A\wedge A\wedge A \big)\,.
\label{H3CS}
\end{equation}
The generators of $\mathcal{G}_{\rm YM}$ satisfy 
$\Tr\,T_IT_J= \tfrac{1}{2}\delta_{IJ}$.

The two-form potential $B^{(2)}$ can be dualized to a six-form $B^{(6)}$
\cite{Cham}.
The action (\ref{2-form-action}) then becomes
\begin{subequations}
\begin{align}
  S_{\textrm{6-form}}
  &= S_{{\rm SG}} + S_{{\rm YM}}\,, \label{action-six-form}
  \\
  S_{\rm{SG}} 
  &= \int \Big( \E^{-\Phi}( R\,{\star\mathbf{1}} + \star\dd\Phi\wedge\dd\Phi ) 
                -\tfrac{1}{2}\E^{\Phi}\,{\star H^{(7)}}\wedge H^{(7)} 
          \Big) \,, \label{action-SG}
  \\
  S_{\rm{YM}}
  &= - \int\Big( 
             \E^{-\Phi} \Tr\,{\star F(A)}\wedge F(A)
             - B^{(6)}\wedge \Tr\,F(A)\wedge F(A)
           \Big) \,, \label{action-YM}
\end{align}
\end{subequations}
where $H^{(7)}=\dd B^{(6)}$. The gauge transformations of $B^{(6)}$
are $\delta B^{(6)}=\dd\Lambda^{(5)}$.

We reduce the action (\ref{action-six-form}) over a compact group $\mathcal{G}$
\cite{SS}. In this section we will limit ourselves to
the supergravity part (\ref{action-SG}). We will analyze the result of this
reduction in Section \ref{analysis1}. In
Section \ref{YM} we will reduce the contribution of the Yang-Mills fields 
(\ref{action-YM}).

The Ansatz for the vielbein is\footnote
{
  From now on ten-dimensional fields
  and indices will be indicated by a hat. The tangent space indices are split up as
  $\hat a = (a, m)$ where $a$ runs over $1,\ldots,4$ and
  $m=1,\ldots,6$, and the curved indices as $\hat
  \mu=(\mu,\alpha)$ where $\mu=1,\ldots,4$ and $\alpha=1,\ldots,6$.
  Further notational matters are discussed in Appendix \ref{conventions}.
}.
\begin{equation}
  \hat e^{a} = e^{a}, \quad
  \hat e^{m} = E_{\alpha}^{m}(\sigma^\alpha - V^\alpha)
               \equiv E_{\alpha}^m\hat{e}^\alpha\,.
\label{e-hat}
\end{equation}
The $V^\alpha=V_{\mu}^{\alpha}\dd x^\mu$ are the Kaluza-Klein gauge fields, and
$E_{\alpha}^{m}$ are scalars which make up the internal metric
$G_{\alpha\beta}=\delta_{mn}E_{\alpha}^{m} E_{\beta}^{n}$.
The $\sigma^\alpha$ are the left-invariant Maurer-Cartan 1-forms
of the internal group\footnote
{
  The volume of $\mathcal{G}$ is normalized as
  $\int_{\mathcal{G}}\sigma^1\wedge\ldots\wedge\sigma^6=1$. 
  The compactness of $\mathcal{G}$ implies that the
  structure constants are traceless, $f^\alpha{}_{\alpha\beta}=0$.
}. 
They satisfy the Maurer-Cartan relation
\begin{equation}
  \dd\sigma^\alpha =
  -\tfrac{1}{2} f^{\alpha}{}_{\beta\gamma}\sigma^{\beta}\wedge\sigma^{\gamma}\,.
\end{equation}
Our parametrization of $\hat e^\alpha$ in (\ref{e-hat}) leads to
\begin{equation}
  \dd\hat e^\alpha 
  = -F(V)^\alpha -f^\alpha{}_{\beta\gamma}\hat e^{\beta}\wedge V^\gamma
    -\tfrac{1}{2}f^\alpha{}_{\beta\gamma}\hat e^\beta\wedge\hat e^\gamma\,.
\end{equation}
$F(V)$, the field strength of the Kaluza-Klein vectors, is given
explicitly by
\begin{equation}
  F(V)^\alpha 
  = \dd V^\alpha + [V,V]^\alpha
  = \dd V^\alpha + \tfrac{1}{2}f^\alpha{}_{\beta\gamma}V^\beta\wedge V^\gamma \,.
\end{equation}
The reduction Ans\"atze for the other fields
are given by\footnote{We will omit the $\wedge$-symbols from now on
to simplify the formulae.}:
\begin{align}
  \hat\Phi 
  &= \phi + \tfrac{1}{2}\ln|\det G_{\alpha\beta}|\,,
  \label{ansatz1}
  \\
  \hat B^{(6)}
  &= \tfrac{1}{2!}B^{(4)}{}_{\alpha_1\alpha_2}\,\hat e^{\alpha_1}\hat e^{\alpha_2}
  +\tfrac{1}{3!} B^{(3)}{}_{\alpha_1\ldots\alpha_3}
  \hat e^{\alpha_1}\cdots\hat e^{\alpha_3}
  +\tfrac{1}{4!} B^{(2)}{}_{\alpha_1\ldots\alpha_4}
  \hat e^{\alpha_1}\cdots \hat e^{\alpha_4}
  \nonumber
  \\
  &\qquad
  +\tfrac{1}{5!} B^{(1)}{}_{\alpha_1\ldots\alpha_5}
  \hat e^{\alpha_1}\cdots\hat e^{\alpha_5}
  +\tfrac{1}{6!} B^{(0)}{}_{\alpha_1\ldots\alpha_6}
  \hat e^{\alpha_1}\cdots\hat e^{\alpha_6}\,.
  \label{ansatz2}
\end{align}
For the $m$-forms $B^{(m)}$ we used the shorthand notation
\begin{equation}
  B^{(m)}{}_{\alpha_1\ldots\alpha_{6-m}}
  =\tfrac{1}{m!} B^{(m)}{}_{a_1\ldots a_m\alpha_1\ldots\alpha_{6-m}} 
  e^{a_1}\cdots  e^{a_m}\,, \quad m=0,1,\ldots,6\,,
\end{equation}
From the Ansatz (\ref{ansatz2}) one obtains the field strength via
$\hat H^{(7)}=\dd\hat B^{(6)}$. By substituting the expression for
$\hat B$ this expands to:
\begin{equation}
  \hat H^{(7)}
  = \sum_{m=1}^{4} \tfrac{1}{m!(7-m)!}
  H^{(m)}{}_{a_1\ldots a_m\alpha_1\ldots\alpha_{7-m}}
  e^{a_1}\cdots e^{a_m} \hat e^{\alpha_1}\cdots \hat e^{\alpha_{7-m}}\,.
\end{equation}

The reduction of the action (\ref{action-SG}) gives the following
four-dimensional result:
\begin{subequations}
\begin{align}
  S 
  &= S_1 + S_2 \,, \label{action-SF}
  \\
  S_1
  &= \int  \E^{-\phi} 
     \Big( 
       R\,{\star\mathbf{1}} + \tfrac{1}{2}{\star\dd\phi}\,\dd\phi
       - \tfrac{1}{2}G_{\alpha\beta}\,{\star F}(V)^\alpha F(V)^\beta
  \nonumber\\
  &    \qquad\qquad\qquad\qquad\qquad
       + \tfrac{1}{4}{\star DG}_{\alpha\beta}DG^{\alpha\beta}
       - V_1(G)\,{\star \mathbf{1}}
     \Big) \,, \label{S1-SF}
  \\
  S_2
  &= -\tfrac{1}{2}\int  \E^\phi\,\det G_{\alpha\beta}
     \Big(
       \tfrac{1}{3!}{\star H}^{(4)\alpha_1\ldots\alpha_3}
       H^{(4)}{}_{\alpha_1\ldots\alpha_3}
       + \tfrac{1}{4!}{\star H}^{(3)\alpha_1\ldots\alpha_4}
       H^{(3)}{}_{\alpha_1\ldots\alpha_4}
  \nonumber\\
  &    \qquad\qquad\qquad\qquad\qquad
       + \tfrac{1}{5!}{\star H}^{(2)\alpha_1\ldots\alpha_5}
       H^{(2)}{}_{\alpha_1\ldots\alpha_5}
       + \tfrac{1}{6!}{\star H}^{(1)\alpha_1\ldots\alpha_6}
       H^{(1)}{}_{\alpha_1\ldots\alpha_6}
     \Big)\,.\label{S2-SF}
\end{align}
\end{subequations}
$D$ is covariant with respect to Scherk-Schwarz gauge transformations.
The contraction in the $H^2$-terms is with $G^{\alpha\beta}$.
The scalar potential $V_1(G)$ due to reduction
of the gravitational sector is given by \cite{SS}
\begin{equation}
  V_1(G) 
  =
  \tfrac{1}{4}G_{\alpha_1\alpha_2}G^{\beta_1\beta_2}G^{\gamma_1\gamma_2}
  f^{\alpha_1}{}_{\beta_1\gamma_1} f^{\alpha_2}{}_{\beta_2\gamma_2}
  + \tfrac{1}{2}G^{\alpha_1\alpha_2}f^{\beta}{}_{\alpha_1\gamma}
  f^{\gamma}{}_{\alpha_2\beta}.
\label{V1}
\end{equation}
The explicit form of the curvatures $H^{(p)}$ is
\begin{subequations}
\label{Hdown}
\begin{align}
  \label{H1down}
  H^{(1)}{}_{\alpha_1\ldots\alpha_6}
  &= DB^{(0)}{}_{\alpha_1\ldots\alpha_6}
  + 15 f^\beta{}_{[\alpha_1\alpha_2}B^{(1)}{}_{\alpha_3\ldots\alpha_6]\beta}\,,
  \\
  \label{H2down}
  H^{(2)}{}_{\alpha_1\ldots\alpha_5}
  &= DB^{(1)}{}_{\alpha_1\ldots\alpha_5}
  + B^{(0)}_{\alpha_1\ldots\alpha_5\beta} F(V){}^{\beta}
  + 10 f^\beta{}_{[\alpha_1\alpha_2}B^{(2)}{}_{\alpha_3\ldots\alpha_5]\beta}\,,
  \\
  \label{H3down}
  H^{(3)}{}_{\alpha_1\ldots\alpha_4}
  &= DB^{(2)}{}_{\alpha_1\ldots\alpha_4}
  + B^{(1)}{}_{\alpha_1\ldots\alpha_4\beta} F(V){}^{\beta}
  +6 f^\beta{}_{[\alpha_1\alpha_2}B^{(3)}{}_{\alpha_3\alpha_4]\beta}\,,
  \\
  \label{H4down}
  H^{(4)}{}_{\alpha_1\ldots\alpha_3}
  &=
  DB^{(3)}{}_{\alpha_1\ldots\alpha_3}
  + B^{(2)}{}_{\alpha_1\ldots\alpha_3\beta}F(V){}^{\beta}
  + 3 f^\beta{}_{[\alpha_1\alpha_2}B^{(4)}{}_{\alpha_3]\beta}\,.
\end{align}
\end{subequations}
The gauge transformations of $B^{(n)}$ are
\begin{equation}
\begin{split}
  \delta B^{(n)}{}_{\alpha_1\ldots\alpha_{6-n}} 
  &= D\Lambda^{(n-1)}{}_{\alpha_1\ldots\alpha_{6-n}}
  - \Lambda^{(n-2)}{}_{\alpha_1\ldots\alpha_{6-n}\beta}F(V)^\beta
  \\
  &\qquad
  -\half(6-n)(5-n) f^{\beta}{}_{[\alpha_1\alpha_2}
  \Lambda^{(n)}{}_{\alpha_3\ldots\alpha_{6-n}]\beta}\,.
  \label{deltaBndown}
\end{split}
\end{equation}

The formulation using lower internal indices is particularly useful
in our cohomology analysis of the Kalb-Ramond degrees of freedom
in Sections \ref{coho} and Appendix \ref{furthercoho}. For other aspects
of the group manifold reduction it is more convenient
to redefine the $p$-form gauge fields by dualizing to
upper internal indices:
\begin{equation}
  B^{(n)}{}_{\alpha_1\ldots\alpha_{6-n}}
  \equiv \tfrac{1}{n!}\,
  \tilde{\ve}_{\alpha_1\ldots\alpha_{6-n}\,\beta_1\ldots\beta_n}
  \tilde{B}^{(n)}{}^{\beta_1\ldots\beta_{n}}\,.
\label{redef}
\end{equation}
The corresponding curvatures are: 
\begin{equation}
\begin{split}
  \tilde{H}^{(n)}{}^{\alpha_1\ldots\alpha_{n-1}} 
  &=
  D\tilde{B}^{(n-1)}{}^{\alpha_1\ldots\alpha_{n-1}}
  + (n-1)(-1)^n \, 
  \tilde{B}^{(n-2)}{}^{[\alpha_1\ldots\alpha_{n-2}}F(V){}^{\alpha_{n-1}]}
  \\ 
  &\qquad
  + \half (n-1)\,f{}^{[\alpha_1}{}_{\beta_1\beta_2}
  \tilde{B}^{(n)}{}^{\alpha_2\ldots\alpha_{n-1}]\beta_1\beta_2}\,.
\label{Hnup}
\end{split}
\end{equation}
The reduced gauge transformations of the $\tilde{B}^{(n)}$ are
\begin{subequations}
\label{deltaBnup}
\begin{align}
  \delta \tilde B^{(4)\alpha_1\ldots\alpha_4} 
  &=
  D\tilde\Lambda^{(3)\alpha_1\ldots\alpha_4}
  -4 F^{[\alpha_1}(V)\tilde\Lambda^{(2)\alpha_2\alpha_3\alpha_4]}
  -2 f^{[\alpha_1}{}_{\beta_1\beta_2}
  \tilde\Lambda^{(4)\alpha_2\alpha_3\alpha_4]\beta_1\beta_2}\,,
  \\
  \delta\tilde B^{(3)\alpha_1\alpha_2\alpha_3}
  &=
  D\tilde\Lambda^{(2)\alpha_1\alpha_2\alpha_3}
  -3 F^{[\alpha_1}(V)\tilde\Lambda^{(1)\alpha_2\alpha_3]}
  - \tfrac{3}{2}f^{[\alpha_1}{}_{\beta_1\beta_2}
  \tilde\Lambda^{(3)\alpha_2\alpha_3]\beta_1\beta_2}\,,
  \\
  \delta\tilde B^{(2)\alpha_1\alpha_2} 
  &=
  D\tilde\Lambda^{(1)\alpha_1\alpha_2}
  -2 F^{[\alpha_1}(V)\tilde\Lambda^{(0)\alpha_2]}
  -  f^{[\alpha_1}{}_{\beta_1\beta_2}
  \tilde\Lambda^{(2)\alpha_2]\beta_1\beta_2}\,,
  \\
  \delta B^{(1)\alpha_1} 
  &=
  D\tilde\Lambda^{(0)\alpha_1}
  -\tfrac{1}{2}f^{\alpha_1}{}_{\beta_1\beta_2}
  \tilde\Lambda^{(1)\beta_1\beta_2}\,,
  \\
  \delta\tilde B^{(0)} 
  &= 0\,.
\end{align}
\end{subequations}
Finally we transform the action (\ref{action-SF}) to the Einstein frame,
and express the result in terms of the curvatures (\ref{Hnup}). We find
\begin{subequations}
\begin{align}
  S &= S_1+S_2\,,
  \label{action-EF}
  \\
  S_1 
  &= 
  \int\Big(
        R\,{\star\mathbf{1}} - \tfrac{1}{2}{\star\dd\phi}\,\dd\phi
        -\tfrac{1}{2}\E^{-\phi}\,G_{\alpha\beta}\,{\star F}(V){}^{\alpha} F(V){}^\beta
  \nonumber\\
  &     \qquad\qquad\qquad
        +\tfrac{1}{4}{\star DG}_{\alpha\beta}DG^{\alpha\beta}
        - \E^\phi\,V_1(G)\,{\star\mathbf{1}}
      \Big)\,,
  \label{S1-EF}
  \\
  S_2
  &= -\tfrac{1}{2}
  \int\Big(
         \tfrac{1}{3!}\E^{-\phi}\,{\star\tilde H}^{(4)\alpha_1\ldots\alpha_3}
         \tilde H^{(4)}{}_{\alpha_1\ldots\alpha_3}
         +
         \tfrac{1}{2}{\star\tilde H}^{(3)\alpha_1\alpha_2}
         \tilde H^{(3)}{}_{\alpha_1\alpha_2}
  \nonumber\\
  &      \qquad\qquad\qquad 
         + \E^\phi\,{\star\tilde H}^{(2)\alpha_1}\tilde H^{(2)}{}_{\alpha_1}
         + \E^{2\phi}\,{\star\tilde H^{(1)}}\tilde H^{(1)}
       \Big)\,,
  \label{S2-EF}
\end{align}
\end{subequations}
where the contraction in the $\tilde H^2$-terms is with $G_{\alpha\beta}$.

\section{Analysis in $D=4$\label{analysis1}}

The massless 3-form fields and the 4-forms that are obtained from the reduction 
of $\hat B^{(6)}$ do not carry physical degrees of freedom. 
The 3-forms however do modify the scalar potential through 
their vacuum expectation values (vevs).
The first part of this section is therefore devoted to an analysis of the
shift symmetries $\delta B^{(n)} \sim  f\Lambda^{(n)}$
in  (\ref{deltaBndown}) in order to determine
the (non)physical components of the $B^{(n)}$
and their masses, which arise through similar terms $H^{(n)}\sim f B^{(n)}$
in the curvatures (\ref{Hdown}).
Section \ref{coho} treats the general case, in Section \ref{example} 
we discuss the example of $\mathcal{G}=SU(2)\times SU(2)$.

\subsection{Cohomology of Scherk-Schwarz gauge transformations\label{coho}}

For a generic field $\Phi$ with $p$ internal indices we have a
local shift symmetry
\begin{equation}
  \delta\Phi_{\alpha_1\ldots\alpha_p} =
  f^\beta{}_{[\alpha_1\alpha_2}\Lambda_{\alpha_3\ldots\alpha_p]\beta}\,,
\label{gauge-away}
\end{equation}
a mass term for this field is constructed from
\begin{equation}
  (m[\Phi])_{\alpha_1\ldots\alpha_{p+1}} =
  f^\beta{}_{[\alpha_1\alpha_2}
    \Phi_{\alpha_3\ldots\alpha_{p+1}]\beta}\,.
\label{mass}
\end{equation}
We associate with each $\Phi$ an element of the set $\Omega^{(p)}$ 
of left-invariant $p$-forms 
on $\mathcal{G}$:
\begin{equation}
  \Phi_{\alpha_1\ldots\alpha_p}\mapsto \Phi^{(p)} =
  \tfrac{1}{p!}\Phi_{\alpha_1\ldots\alpha_p}\sigma^{\alpha_1}\ldots\sigma^{\alpha_p}\,.
\end{equation}
Given the exterior derivative $\dd_{p}:\Omega^{(p)}\rightarrow \Omega^{(p+1)}$,
we define its image $C^{(p)}\equiv \textrm{Im}\,(\dd_{p-1})$, its kernel
$Z^{(p)}=\textrm{Ker}\,(\dd_{p})$ and the quotient $H^{(p)}=Z^{(p)}/C^{(p)}$.
The equations (\ref{gauge-away}), (\ref{mass}) are equivalent to a 
cohomology problem for the left-invariant forms:
\begin{equation}
  \delta\Phi^{(p)} = \dd\Lambda^{(p-1)},\ m[\Phi^{(p)}]=\dd\Phi^{(p)}\,.
\end{equation}
A well-known theorem by Chevalley and Eilenberg \cite{chev} tells us that
this cohomology problem is equivalent to the usual de Rham cohomology.
In particular
\begin{equation}
  \textrm{dim}\, H^{(p)}= b_p,
\end{equation}
where $b_p$ is the $p$th Betti number of $\mathcal{G}$.
Since $d_p$ is a homomorphism we have
\begin{equation}
  C^{(p)}\cong \Omega^{(p-1)}/Z^{(p-1)}\,,
\end{equation}
giving us the recurrence relation
\begin{equation}
  \textrm{dim} \,C^{(p+1)} = \textrm{dim}\,\Omega^{(p)} -
  \textrm{dim}\,C^{(p)}-b_p = \left(\begin{array}{c} 6 \\ p
  \\\end{array}\right) - b_p - \textrm{dim}\,C^{(p)}
\end{equation}
This recurrence relation can be solved starting from
dim$\,C^{(0)}=0$:
\begin{equation}
\begin{split}
  \textrm{dim}\,C^{(0)} &= 0 \\
  \textrm{dim}\,C^{(1)} &= 1-b_0 \\
  \textrm{dim}\,C^{(2)} &= 5-b_1+b_0 \\
  \textrm{dim}\,C^{(3)} &= 10-b_2+b_1-b_0 \\
  \textrm{dim}\,C^{(4)} &= 10 -b_3+b_2-b_1+b_0 \\
  \textrm{dim}\,C^{(5)} &= 5 -b_4+b_3-b_2+b_1-b_0 = 5-\chi(\mathcal{G})+b_6-b_5 \\
  \textrm{dim}\,C^{(6)} &= 1-b_5+b_4-b_3+b_2-b_1+b_0 = 1+\chi(\mathcal{G})-b_6,
\end{split}
\label{dimlist}
\end{equation}
where $\chi(\mathcal{G})$ is the Euler characteristic
\begin{equation}
  \chi(\mathcal{G})=\sum_{r=0}^{6}(-1)^rb_r.
\end{equation}
We draw the following conclusions:
\begin{itemize}
  \item[--] The mass terms are invariant under the shift symmetries.
  \item[--] A field can be gauged away if $\Phi^{(p)} \in C^{(p)}$.
            These fields do not have mass terms.
  \item[--] If and only if a field $\Phi^{(p)}$ has a mass term, there is a
  $\Phi^{(p+1)}$ that can be gauged away, i.e. $\Phi^{(p+1)}$ is `eaten'
  by $\Phi^{(p)}$.
  \item[--] Physical massless fields are elements of $H^{(p)}$, 
            the number of these fields is $b_p$.
  \item[--] The number of massive fields in $\Omega^{(p)}$ is ${\rm dim}\,C^{(p+1)}$.
\end{itemize}

In this context the fields $B^{(n)}$ in four dimensions are interpreted
as elements of $\Omega^{(6-n)}$, for $n=0,\ldots,4$.
The implications for these fields
are presented in Table \ref{table1}. The total number of physical
degrees of freedom in four dimensions is obtained by taking,
for each row of Table \ref{table1}, the product of the dimension in the third,
and the number of degrees of freedom in the fourth column, and
by summing these products.
Using (\ref{dimlist}) one finds that this
sum is $28+\chi(\mathcal{G})$. For a compact connected group,
$\chi(\mathcal{G})=0$ (see e.g. \cite{frankel}); we thus recover the 28 degrees
of freedom of the ten-dimensional $\hat B^{(6)}$.

\begin{table}
\begin{center}
\label{table1}
  \begin{tabular}[tbp]{|c|l|c|c|}
    \hline field & features& dimension   & $\#$ DOF \\
    \hline
    \ $B^{(4)}$  &gauge    & ${\rm dim}\,C^{(2)}$ & $0$ \\
    \            &massive  & ${\rm dim}\,C^{(3)}$ & $0$ \\
    \            &massless & $b_2$ & $0$ \\
    \hline
    \ $B^{(3)}$  &gauge    & ${\rm dim}\,C^{(3)}$ & $0$ \\
    \            &massive  & ${\rm dim}\,C^{(4)}$ & $1$ \\
    \            &massless & $b_3$                & $0$\\
    \hline
		\ $B^{(2)}$  &gauge    & ${\rm dim}\,C^{(4)}$ & $0$\\
		\            &massive  & ${\rm dim}\,C^{(5)}$ & $3$ \\
		\            &massless & $b_4$                & $1$ \\
		\hline
		\ $B^{(1)}$  &gauge    & ${\rm dim}\,C^{(5)}$ & $0$ \\
		\            &massive  & ${\rm dim}\,C^{(6)}$ & $3$ \\
		\            &massless & $b_5$                & $2$ \\
		\hline
		\ $B^{(0)}$  &gauge    & ${\rm dim}\,C^{(6)}$ & 0\\
		\            &massive  & $0$                  & $1$ \\
		\            &massless & $b_6$                & $1$ \\
		\hline
  \end{tabular}
\caption{{\small Result of the analysis of the
  Scherk-Schwarz gauge transformations of the Kalb-Ramond fields
  $B^{(n)}$. Each $B^{(n)}$ splits in three parts: gauge degrees
  of freedom, massive and massless components. The third column
  shows how the dimension of the various spaces $C^{(p)}$ and
  $H^{(p)}$ determines the number of components of $B^{(n)}$.
  DOF indicates the
  number of spacetime degrees of freedom for each choice
  of the $6-n$ internal indices. Hence the total number of degrees of
  freedom per field is obtained by multiplying horizontally.}}
\end{center}
\end{table}

We conclude from the above cohomology analysis that we can give a vev
to a $B^{(3)}$ if $b_3\neq 0$.
This is always the case, since for any compact nonabelian group there is
a nonzero harmonic 3-form
\begin{equation}
  \Omega_3=\textrm{Tr}(g^{-1}dg\wedge g^{-1}dg\wedge g^{-1}dg),\quad
  g\in \mathcal{G} \label{stellingCartan}\,.
\end{equation}

In the case of $B^{(4)}$ the equation of motion is algebraic, it sets,
in the absence of Yang-Mills fields, certain components of
$H^{(4)}$ to zero. The $b_2$ massless components in Table \ref{table1} do not
appear in the action at all, and never give rise to fluxes.

If $\mathcal{G}$ is semi-simple, $b_1=b_5=0$.
Combining this with $b_0=b_6=1$ and $\chi(\mathcal{G})=0$, we see that
all vectors $B^{(1)}$ can be gauged away. 

Explicit examples of the analysis for specific group manifolds are
given in the following subsection, and in Appendix \ref{furthercoho}.

\subsection{Analysis for $SU(2)\times SU(2)$\label{example}}

In this subsection we will work out the example of the gauge group
$SU(2)\times SU(2)$. We first give the results following
the methods of Section \ref{coho}. Then we rederive the
results by analyzing the Scherk-Schwarz gauge transformations explicitly.

For $SU(2)\times SU(2)$ the Betti numbers are given by
\begin{equation}
  G=SU(2)\times SU(2)\Rightarrow (b_0,\ldots,b_6) = (1,0,0,2,0,0,1)
\end{equation}
The resulting degrees of freedom are given in Table \ref{table2}.
$B^{(2)}$ and $B^{(3)}$ give only massive degrees of freedom, which
are therefore stabilized. The single massless degree of freedom 
corresponds to the axion.

\begin{table}
\begin{center}
\label{table2}
  \begin{tabular}[tbp]{|c|l|c|c|c|}
		\hline field & features & dimension & \# DOF & total DOF\\
		\hline
		\ $B^{(4)}$ &  gauge  & $6$ &    0  & 0 \\
		\ &  massive          & $9$ &    0  & 0 \\
		\ &  massless         & $0$ &    0  & 0 \\
		\hline
		\ $ B^{(3)}$ &  gauge   & $9$ &  0  & 0\\
		\ &  massive            & $9$ &  1  & 9\\
		\ & massless            & $2$ &  0  & 0\\
		\hline
		\ $B^{(2)}$ &  gauge    & $9$ &  0  & 0\\
		\ &  massive            & $6$ &  3  &18 \\
		\ &  massless           & $0$ &  1  & 0 \\
		\hline
		\ $B^{(1)}$ &  gauge    & $6$ &  0  & 0 \\
		\ &  massive            & $0$ &  3  & 0  \\
		\ &  massless           & $0$ &  2  & 0  \\
		\hline
		\ $B^{(0)}$ & gauge     & $0$ &  0  & 0\\
		\ & massive             & $0$ &  1  & 0\\
		\ & massless            & $1$ &  1  & 1\\
		\hline
  \end{tabular}
\caption{{\small
  Result of the analysis of the
  Scherk-Schwarz gauge transformations of the Kalb-Ramond fields
  $B^{(n)}$ for the group $SU(2)\times SU(2)$, in the reduction
  of $B^{(6)}$ from 10 to 4 dimensions. See the caption
  of Table \ref{table1} for more details.
  In the last column we indicate
  the total numbers of degrees of freedom.
}}
\end{center}
\end{table}

In our explicit analysis we indicate the six internal
directions by indices $a,b,\ldots=1,2,3$
and $\bar{a},\bar{b},\ldots=4,5,6$, with one $SU(2)$ gauging the $a,b,\ldots$, and the
other the $\bar{a},\bar{b},\ldots$ directions. In this analysis,
which uses (\ref{Hnup},\ \ref{deltaBnup}), we first
investigate which field components of $\tilde B^{(n)}$ can be gauged away by the
shift transformations involving $\tilde\Lambda^{(n)}$, and then look at the impact of
this process on the equations of motion.

To start, all six vectors $\tilde B^{(1)}$ can be gauged away by the transformations
\begin{equation}
  \delta\tilde B^{(1)}{}^{a} = -\half f^{a}{}_{cd}\tilde\Lambda^{(1)}{}^{cd}\,,\quad
  \delta\tilde B^{(1)}{}^{\bar{a}} = -\half f^{\bar{a}}{}_{\bar{c}\bar{d}}
    \tilde\Lambda^{(1)}{}^{\bar{c}\bar{d}}\,.
\end{equation}
Note that this does not involve the parameters $\Lambda^{a\bar{a}}$, which therefore
remain free. The vectors which we gauge away will contribute to
masses for $\tilde B^{(2)}{}^{ab}$ and $\tilde B^{(2)}{}^{\bar{a}\bar{b}}$.
These mass terms are invariant under the shift
transformations of $\tilde B^{(2)}$, due to the Jacobi identity for the structure constants.
The remaining fields $\tilde B^{(2)}{}^{a\bar{a}}$ transform as
\begin{equation}
  \delta\tilde B^{(2)}{}^{a\bar{a}} = - \big(
   f^{a}{}_{cd}\tilde\Lambda^{(2)}{}^{\bar{a}cd} -
   f^{\bar{a}}{}_{\bar{c}\bar{d}}\tilde\Lambda^{(2)}{}^{a\bar{c}\bar{d}} \big)\,.
\end{equation}
This implies that these components of $B^{(2)}$ can be gauged away. The
components
\begin{equation}
  \tilde B^{(3)}_A{}^{a\bar{a}} = f^{a}{}_{cd}\tilde B^{(3)}{}^{\bar{a}cd} -
    f^{\bar{a}}{}_{\bar{c}\bar{d}}\tilde B^{(3)}{}^{a\bar{c}\bar{d}}
\label{B3A}
\end{equation}
then become massive. All this implies that 6 vectors are gauged away,
there are 6 massive 2-forms corresponding to 18 degrees of freedom, and 9
massive 3-forms, which have 9 degrees of freedom. Together with the
axion $\tilde B^{(0)}$ we find the 28 degrees of freedom of $\hat B^{(6)}$.

For $\tilde B^{(3)}$ the remaining components are $\tilde B^{(3)}{}^{abc}$ and
$\tilde B^{(3)}{}^{\bar{a}\bar{b}\bar{c}}$, which are both gauge invariant under
$SU(2)\times SU(2)$ and cannot be gauged away, and the symmetric combination
\begin{equation}
  \tilde B^{(3)}_S{}^{a\bar{a}} = f^{a}{}_{cd}\tilde B^{(3)}{}^{\bar{a}cd} +
    f^{\bar{a}}{}_{\bar{c}\bar{d}}\tilde B^{(3)}{}^{a\bar{c}\bar{d}}\,.
\label{B3S}
\end{equation}
Under the shifts of $\tilde B^{(3)}$ this transforms as
\begin{equation}
  \delta\tilde B^{(3)}_S{}^{a\bar{a}} = -3
   f^{a}{}_{cd}f^{\bar{a}}{}_{\bar{c}\bar{d}} 
   \tilde\Lambda^{(3)}{}^{cd\bar{c}\bar{d}}\,,
\end{equation}
so that these symmetric components can be gauged away. This implies that the terms
involving $\tilde B^{(4)}{}^{ab\bar{a}\bar{b}}$ remain in the action. In fact, they are
(for the chosen gauge group) the only components of $\tilde B^{(4)}$ which appear in the
action in the first place.

The other components of $\tilde B^{(3)}$ and those of $\tilde B^{(4)}$ can be treated as
follows. The components $\tilde B^{(3)}{}^{abc}$ and 
$\tilde B^{(3)}{}^{\bar{a}\bar{b}\bar{c}}$
are gauge invariant, and they do not appear in $\tilde H^{(3)}$. Their field
equation is therefore simply:
\begin{equation}
  D\big(  \E^{-\phi}\,
  {\star\tilde H}^{(4)}{}_{abc}\big) = 0
  \,,\qquad
  D\big( \E^{-\phi}\,
  {\star\tilde H}^{(4)}{}_{\bar{a}\bar{b}\bar{c}}\big) = 0\,,
\label{H4abc}
\end{equation}
where internal indices have been lowered with the internal metric. We conclude
that these components of $\tilde H^{(4)}$ give two independent constants: they are
the $\tilde B^{(3)}$ vevs discussed in Section \ref{coho}. 
These vevs are precisely those given by
(\ref{stellingCartan}) for the two $SU(2)$-factors:
\begin{equation}
  \tilde H_{(4)}^{{\rm vev}} \sim
  m_1\,\varepsilon_{abc}\,\sigma^a\sigma^b\sigma^c
  +m_2\,\varepsilon_{\bar a\bar b\bar c}\,\sigma^{\bar a}\sigma^{\bar b}\sigma^{\bar c}\,.
\end{equation}
Of course in this analysis we ignore the interaction with other fields, we
will come back to the complete equations of motion in Section \ref{potential}.

The equation of motion of the remaining $\tilde B^{(4)}$ components is
\begin{equation}
  \tilde H^{(4)}{}_{a\bar{c}\bar{d}} f^{{a}}{}_{cd} +
  \tilde H^{(4)}{}_{\bar{a}cd}f^{\bar{a}}{}_{\bar{c}\bar{d}} = 0\,,
\end{equation}
thus putting the field strengths of the unphysical
symmetric combination $\tilde B^{(3)}_S{}^{a\bar{a}}$ to zero.

Therefore we see that the propagating degrees of freedom in this case come
from  $\tilde B^{(2)}{}^{ab}$ and $\tilde B^{(2)}{}^{\bar{a}\bar{b}}$, and the
$\tilde B^{(3)}_A{}^{a\bar{a}}$ (\ref{B3A}). This of course agrees with
the cohomology analysis in Section \ref{coho}.
The nonpropagating degrees of freedom
$\tilde B^{(3)}{}^{abc}$ and $\tilde B^{(3)}{}^{\bar{a}\bar{b}\bar{c}}$ lead to constant
components of $\tilde H^{(4)}$ which contribute to the scalar potential.

\section{Coupling to Yang-Mills multiplets\label{YM}}

In this section we reduce the Yang-Mills sector (\ref{action-YM}). 
The Ansatz for the ten-dimensional Yang-Mills field is
\begin{equation}
  \hat A^I= A^{I}{}_{a}e^a + A^{I}{}_{\alpha}\hat e^{\alpha}\,.
\label{ansatz-YM}
\end{equation}
The field strength can then be obtained from
$\hat F(\hat A)^I = d\hat A^I + (\hat A\wedge \hat A)^I$
with the result\footnote
{
  In the following $\bar D$
  stands for a covariant derivative which is covariant with
  respect to both the Scherk-Schwarz and Yang-Mills gauge transformations.
  Again we omit the $\wedge$-symbols. It should be
  kept in mind that $A^I$ is the one-form corresponding to the gauge vector,
  while $A^I{}_\alpha$ are scalar fields. The internal indices
  $\alpha,\beta,\ldots$ are therefore always written explicitly.
}:
\begin{equation}
  \hat F(\hat A)^I  = G^I + \bar D A^{I}{}_{\alpha} e^\alpha
  +\tfrac{1}{2}\mathcal{F}^{I}{}_{\alpha_1\alpha_2}\,\hat e^{\alpha_1}
  \hat e^{\alpha_2}\,,
\end{equation}
where we use the following definitions:
\begin{subequations}
\begin{align}
  F(A)^I 
  &= \dd A^I +\tfrac{1}{2}f^I{}_{JK}A^J A^K\,,
  \\
  G^I 
  &= F(A)^I - A^{I}{}_{\alpha}F^{\alpha}(V)\,,
  \\
  \bar D A^{I}{}_{\alpha}
  &= \dd A^{I}{}_{\alpha}
  +f^\beta{}_{\alpha\gamma}V^\gamma A^{I}{}_{\beta}
  +f^I{}_{JK} A^J A^{K}{}_{\alpha}\,,
  \\
  \mathcal{F}^{I}{}_{\alpha_1\alpha_2} 
  &=
  f^I{}_{JK}A^{J}{}_{\alpha_1}A^{K}{}_{\alpha_2} -
  A^{I}{}_{\gamma}f^\gamma{}_{\alpha_1\alpha_2}\,.
\end{align}
\end{subequations}
The ten-dimensional Chern-Simons form $\hat{C}$ 
can be rewritten as
\begin{equation}
  \hat C=
  \textrm{Tr}\, (\hat A\, \dd\hat A +\tfrac{2}{3}\hat
  A\, \hat A\, \hat A) = \tfrac{1}{2}\hat A^I \, \hat F(\hat A)^I
  -\tfrac{1}{12}f^I{}_{JK}\hat A^I \,\hat A^J \,\hat A^K\,.
\end{equation}
Substitution of the Ansatz (\ref{ansatz-YM}) gives
\begin{equation}
  \hat C = C^{(3)}+C^{(2)}{}_{\alpha} \hat e^\alpha
  +\tfrac{1}{2!}C^{(1)}{}_{\alpha_1\alpha_2}
  \hat e^{\alpha_1}\hat e^{\alpha_2}
  +\tfrac{1}{3!}C^{(0)}{}_{\alpha_1\alpha_2\alpha_3}
    \hat e^{\alpha_1}\hat e^{\alpha_2}\hat e^{\alpha_3} \,,
\label{CS10}
\end{equation}
where
\begin{subequations}
\begin{align}
  C^{(3)} 
  &= \tfrac{1}{2}(A^I G^I - \tfrac{1}{6}f^I{}_{JK}A^I A^J A^K)\,,
  \\
  C^{(2)}{}_\alpha 
  &=
  \tfrac{1}{2}(A^I\bar D A^{I}{}_{\alpha}+A^{I}{}_{\alpha} G^I
  -\tfrac{1}{2}f^I{}_{JK}A^I A^J A^{K}{}_{\alpha})\,,
  \\
  C^{(1)}{}_{\alpha_1\alpha_2}
  &=
  -\tfrac{1}{2}A^I A^{I}{}_{\gamma}f^\gamma{}_{\alpha_1\alpha_2} -
  A^{I}{}_{[\alpha_1}\bar D A^{I}{}_{\alpha_2]}\,,
  \\
  C^{(0)}{}_{\alpha_1\alpha_2\alpha_3} 
  &=
  f^I{}_{JK} A^{I}{}_{\alpha_1}A^{J}{}_{\alpha_2}A^{K}{}_{\alpha_3}
  -\tfrac{3}{2} A^I{}_{[\alpha_1} f^\delta{}_{\alpha_2\alpha_3]}A^{I}{}_{\delta}\,.
\end{align}
\end{subequations}
The reduction of the ten-dimensional action
(\ref{action-YM}) is facilitated by using
$\dd\hat C  = \tfrac{1}{2}\hat F^I\hat F^I$
to write
\begin{equation}
  {\cal L}_{{\rm CS}} = \hat B^{(6)}\wedge \textrm{Tr}\,(\hat F \wedge \hat F )=
  - \hat H^{(7)}\wedge \hat C^{(3)} + \textrm{ total derivative.}
\end{equation}
After some algebra one obtains the following contribution to
the four-dimensional action:
\begin{equation}
  \mathcal{L}_{{\rm CS}} =
  -\tilde H^{(1)}C^{(3)}
  - \tilde H^{(2)\alpha}C^{(2)}{}_{\alpha}
  - \tfrac{1}{2!}\tilde H^{(3)\alpha_1\alpha_2}C^{(1)}{}_{\alpha_1\alpha_2}
  - \tfrac{1}{3!}\tilde H^{(4)\alpha_1\alpha_2\alpha_3}
    C^{(0)}{}_{\alpha_1\alpha_2\alpha_3}\,,
\end{equation}
which after a partial integration gives
\begin{equation}
\begin{split}
  \mathcal{L}_{{\rm CS}} 
  &= +\tfrac{1}{2} \tilde{B}^{(0)} G^I G^I 
  - \tilde B^{(1)}{}^\alpha G^I\bar D A^{I}{}_{\alpha}
  + \tfrac{1}{2}( G^I\mathcal{F}^I{}_{\alpha_1\alpha_2} 
                  - \bar D A^{I}{}_{\alpha_1}\bar D A^{I}{}_{\alpha_2} ) 
                \tilde B^{(2)}{}^{\alpha_1\alpha_2}
  \\
  &\qquad\qquad
  - \tfrac{1}{2}\tilde B^{(3)\alpha_1\alpha_2\alpha_3}\bar D
    A^{I}{}_{\alpha_1}\mathcal{F}^{I}{}_{\alpha_2\alpha_3}
  + \tfrac{1}{8}\tilde B^{(4)\alpha_1\ldots\alpha_4}
    \mathcal{F}^{I}{}_{\alpha_1\alpha_2}
    \mathcal{F}^{I}{}_{\alpha_3\alpha_4}\,. \label{bff-term-reductie}
\end{split}
\end{equation}

We now give the complete four-dimensional Lagrangian
with Yang-Mills fields, in the Einstein frame:
\begin{subequations}
\label{finalL}
\begin{align}
  \mathcal{L}
  &= \mathcal{L}_1+\mathcal{L}_2+\mathcal{L}_3+\mathcal{L}_4\,,
  \\
  \mathcal{L}_1 
  &= \sqrt{-g}\,
  \Big( R - \tfrac{1}{2}\partial_\mu \phi\partial^\mu \phi
    - \tfrac{1}{4}\E^{-\phi}\, F(V)_{\mu\nu}{}^\alpha 
    G_{\alpha\beta}F(V)^{\mu\nu\,\beta}
  \nonumber
  \\
  &\qquad\qquad\qquad
    + \tfrac{1}{4}D_\mu G_{\alpha\beta} D^\mu G^{\alpha\beta}
    - \E^{\phi}\,V_1(G)
  \Big)\,,
  \label{finalL1}
  \\
  \mathcal{L}_2 
  &=
  -\sqrt{-g}\,
  \Big(
    \tfrac{1}{2}\E^{2\phi}\,\partial_\mu\tilde{B}^{(0)}\partial^\mu\tilde{B}^{(0)}
    +\tfrac{1}{4}\E^\phi\,
     \tilde H^{(2)}{}_{\mu\nu}{}^{\alpha}
     \tilde H^{(2)}{}^{\mu\nu\,\beta}
     G_{\alpha\beta}
  \nonumber
  \\
  &\qquad\qquad\qquad
    +\tfrac{1}{4!}
     \tilde H^{(3)}{}_{\mu\nu\lambda}{}^{\alpha_1\alpha_2}
     \tilde H^{(3)}{}^{\mu\nu\lambda}{}^{\beta_1\beta_2}
     G_{\alpha_1\beta_1}G_{\alpha_2\beta_2}
  \nonumber
  \\
  &\qquad\qquad\qquad
    +\tfrac{1}{2\cdot 3!4!}\E^{-\phi}
     \tilde H^{(4)}{}_{\mu\nu\lambda\rho}{}^{\alpha_1\alpha_2\alpha_3}
     \tilde H^{(4)}{}^{\mu\nu\lambda\rho}{}^{\beta_1\beta_2\beta_3}
     G_{\alpha_1\beta_1}G_{\alpha_2\beta_2}G_{\alpha_3\beta_3}
  \Big)\,,
  \label{finalL2}
  \\
  \mathcal{L}_3
  &=
  -\sqrt{-g}\,\E^{-\phi}
  \Big(
    \tfrac{1}{4}G^{I}{}_{\mu\nu}G^{I\,\mu\nu}
    +\tfrac{1}{2}\E^{\phi}\,\bar D_\mu A^{I}{}_{\alpha}
     \bar D^\mu A^{I}{}_{\beta} G^{\alpha\beta}
  \nonumber
  \\
  &\qquad\qquad\qquad
    +\tfrac{1}{4}\E^{2\phi}\,\mathcal{F}^{I}{}_{\alpha_1\alpha_2}
     \mathcal{F}^{I}{}_{\beta_1\beta_2}
     G^{\alpha_1\beta_1}G^{\alpha_2\beta_2}
  \Big)\,,
  \label{finalL3}
  \\
  \mathcal{L}_4 
  &= -\tilde\epsilon^{\mu\nu\lambda\rho}
  \Big( 
    \tfrac{1}{8}\tilde{B}^{(0)} G^{I}{}_{\mu\nu} G^{I}{}_{\lambda\rho}
    -\tfrac{1}{2}\tilde B^{(1)}{}_\mu{}^{\alpha}\,
     G^{I}{}_{\nu\lambda} \bar D_{\rho}A^{I}{}_{\alpha}
  \nonumber
  \\
  &\qquad\qquad\qquad
    +\tfrac{1}{4}\tilde B^{(2)}{}_{\lambda\rho}{}^{\alpha\beta}
     \big( 
       \tfrac{1}{2}G^{I}{}_{\mu\nu}\mathcal{F}^I{}_{\alpha\beta}
       - \bar D_\mu A^I{}_{\alpha}\bar D_\nu A^I{}_{\beta}
     \big)
  \nonumber
  \\
  &\qquad\qquad\qquad
    -\tfrac{1}{2\cdot 3!}\tilde B^{(3)}{}_{\mu\nu\lambda}{}^{\alpha\beta\gamma}
     \bar D_\rho A^I{}_{\alpha}\mathcal{F}^I{}_{\beta\gamma}
  +\tfrac{1}{8\cdot 4!}
     \tilde B^{(4)}{}_{\mu\nu\lambda\rho}{}^{\alpha\beta\gamma\delta}
     \mathcal{F}^I{}_{\alpha\beta} \mathcal{F}^I{}_{\gamma\delta}
  \Big)\,.
\label{finalL4}
\end{align}
\end{subequations}

\section{Scalar potential and extrema\label{potential}}

In this section we will analyze the scalar potential. The potential consists
of $V_1(G)$ (\ref{V1}) and terms involving the Yang-Mills scalars
-- the last term in (\ref{finalL3}).
In addition there are contributions from the vevs which can be introduced using 
$\tilde H^{(4)}$
-- the last term in (\ref{finalL2}), see sections \ref{coho} and \ref{example}.
Note that in introducing these vevs we solve the equation of motion 
for certain components of $\tilde B^{(3)}$. 
These solutions should not be substituted back into the action,
and therefore we will determine the potential from the equations of motion.
To show that the same modified scalar potential appears in different equations
of motion we work this out using an appropriate subset of fields from
(\ref{finalL}), namely the metric, the dilaton $\phi$, the scalars
$G_{\alpha\beta}$ and $\tilde B^{(3)}$. Later in this section we will investigate
the complete equations of motion.

We therefore  consider the following action (see also \cite{HT}):
\begin{equation}
\begin{split}
  S
  &=\int\dd^4 x\,\sqrt{-g}
  \Big(
    R -\tfrac{1}{2}\partial_\mu \phi\partial^\mu \phi
    +\tfrac{1}{4}\partial_\mu G_{\alpha\beta} \partial^\mu G^{\alpha\beta}
    -\E^{\phi}\, V_1(G)
  \nonumber
  \\
  &\qquad\qquad
    -\tfrac{1}{2\cdot 3!4!}\E^{-\phi}\,
     G_{\alpha_1\beta_1}G_{\alpha_2\beta_2}G_{\alpha_3\beta_3}
     \tilde H^{(4)}{}_{\mu_1\ldots\mu_4}{}^{\alpha_1\alpha_2\alpha_3}
     \tilde H^{(4)}{}^{\beta_1\beta_2\beta_3\mu_1\ldots\mu_4}
  \Big)\,.
\end{split}
\end{equation}
The equations of motion read:
\begin{subequations}
\begin{align}
  R_{\mu\nu}-\tfrac{1}{2}g_{\mu\nu}R
  &=
  - \tfrac{1}{4}\textrm{tr}\,\partial_\mu G\partial_\nu G^{-1}
  + \tfrac{1}{2}\partial_\mu \phi \partial_\nu \phi
  + g_{\mu\nu}\big( \tfrac{1}{8}\textrm{tr}\,\partial G \partial G^{-1}
                    -\tfrac{1}{4}\partial\phi \partial\phi \big)
  \nonumber
  \\
  & \qquad\qquad
  + \tfrac{1}{2\cdot 3!3!} \E^{-\phi}
    \big( (\tilde H^2)_{\mu\nu}- \tfrac{1}{8} g_{\mu\nu}\tilde H^2 \big)
  - \tfrac{1}{2}g_{\mu\nu}\E^{\phi}\,V_1(G)\,,
  \label{eom-g}
  \\
  0
  &= \Box \phi - \E^\phi V_1(G)+\tfrac{1}{2\cdot 3!4!}\E^{-\phi}\tilde H^2\,,
  \label{eom-phi}
  \\
  0
  &= \tfrac{1}{4}\big( G^{-1}(\Box G) G^{-1}-\Box G^{-1} \big)^{\alpha\beta}
  -\tfrac{1}{2\cdot 2!4!} \E^{-\phi}(\tilde H^2)^{\alpha\beta}
  -\E^\phi\frac{\partial}{\partial G_{\alpha\beta}}V_1(G)\,,
  \label{eom-G}
  \\
  0
  &= \nabla_{\mu_1}
  \big( \E^{-\phi}\,\tilde H^{(4)\mu_1\ldots\mu_4\,\alpha_1\alpha_2\alpha_3}
        G_{\alpha_1\beta_1} G_{\alpha_2\beta_2} G_{\alpha_3\beta_3}
  \big)\,.
  \label{eom-H4}
\end{align}
\end{subequations}
We have used the abbreviations
\begin{equation}\label{H2defs}
\begin{split}
  \tilde H^2 
  &=
  G_{\alpha_1\beta_1}G_{\alpha_2\beta_2}G_{\alpha_3\beta_3}
  \tilde H^{(4)}{}_{\mu\nu\lambda\rho}{}^{\alpha_1\alpha_2\alpha_3}
  \tilde H^{(4)}{}^{\mu\nu\lambda\rho}{}^{\beta_1\beta_2\beta_3}\,,
  \\
  (\tilde H^2)_{\mu\nu}
  &= 
  G_{\alpha_1\beta_1}G_{\alpha_2\beta_2}G_{\alpha_3\beta_3}
  \tilde H^{(4)}{}_{\mu\sigma\lambda\rho}{}^{\alpha_1\alpha_2\alpha_3}
  \tilde H^{(4)}{}_{\nu}{}^{\sigma\lambda\rho}{}^{\beta_1\beta_2\beta_3}\,,
  \\
  (\tilde H^2)^{\alpha\beta}
  &=
  G_{\gamma_1\delta_1}G_{\gamma_2\delta_2}
  \tilde H^{(4)}{}_{\mu\nu\lambda\rho}{}^{\alpha\gamma_1\gamma_2}
  \tilde H^{(4)}{}^{\mu\nu\lambda\rho}{}^{\beta\delta_1\delta_2}\,.
\end{split}
\end{equation}
We now define ${\cal H}_{\alpha\beta\gamma}$ by
\begin{equation}
  G_{\alpha_1\beta_1} G_{\alpha_2\beta_2}G_{\alpha_3\beta_3}
  \tilde H^{(4)}{}_{\mu_1\mu_2\mu_3\mu_4}{}^{\alpha_1\alpha_2\alpha_3}
  =
  \E^{\phi}\,\varepsilon_{\mu_1\mu_2\mu_3\mu_4}\,
  \mathcal{H}_{\beta_1\beta_2\beta_3}
  \,.
\label{redefH4}
\end{equation}
The equation of motion (\ref{eom-H4}) implies that
$\mathcal{H}_{\alpha\beta\gamma}$ is constant.
The equations (\ref{eom-g}-\ref{eom-G}) can now be
expressed in terms of
\begin{equation}
  \mathcal{H}^2 \equiv \mathcal{H}_{\alpha\beta\gamma}
  \mathcal{H}_{\alpha'\beta'\gamma'}
  G^{\alpha\alpha'}G^{\beta\beta'}G^{\gamma\gamma'}\,.
\end{equation}
The result is
\begin{subequations}
\begin{align}
  R_{\mu\nu}-\tfrac{1}{2}g_{\mu\nu}R 
  &=
  - \tfrac{1}{4}\textrm{tr}\,\partial_\mu G\partial_\nu G^{-1}
  + \tfrac{1}{2}\partial_\mu \phi \partial_\nu \phi
  + g_{\mu\nu}\big( \tfrac{1}{8}\textrm{tr}\,\partial G \partial G^{-1}
                    - \tfrac{1}{4}\partial\phi\partial\phi \big)
  \nonumber                    
  \\
  & \qquad\qquad
  - \tfrac{1}{4\cdot 3!} g_{\mu\nu}\,\E^{\phi}\,\mathcal{H}^2
  - \tfrac{1}{2}g_{\mu\nu}\,\E^{\phi}\,V_1(G)\,,
  \\
  0
  &=
  \Box \phi - \E^\phi\, V_1(G)
  - \tfrac{1}{2\cdot 3!}\E^{\phi}\,\mathcal{H}^2\,,
  \\
  0
  &= 
  \tfrac{1}{4}\big( G^{-1}(\Box G) G^{-1}-\Box G^{-1} \big)^{\alpha\beta}
  - \E^\phi\frac{\partial}{\partial G_{\alpha\beta}}
  \Big( \tfrac{1}{2\cdot 3!}\mathcal{H}^2 + V_1(G) \Big)\,.
\end{align}
\end{subequations}
Here we use that
\begin{equation}
  (\tilde H)^{2\alpha\beta} = \tfrac{4!}{3}\,\E^{2\phi}
  \frac{\partial}{\partial G_{\alpha\beta}} \mathcal{H}^2\,.
\end{equation}
Hence the potential has effectively changed to
\begin{equation}
  V_{{\rm eff}}(G)
  =\E^\phi \Big( V_1(G)+\tfrac{1}{2\cdot 3!} \mathcal{H}^2 \Big)\,.
\label{Vflux}
\end{equation}
The complete scalar potential also includes the contributions from the
Yang-Mills fields.  But also the equations
of motion of $\tilde B^{(3)}$ and  $\tilde B^{(4)}$  will be modified  by Yang-Mills
contributions, and in case of $\tilde B^{(3)}$, the equation of motion
contains additional terms due to the appearance of $\tilde B^{(3)}$ in
$\tilde H^{(3)}$.

Let us continue the analysis for the specific case
$\mathcal{G}=SU(2)\times SU(2)$. Then the $\tilde B^{(3)}$ contributions in 
$\tilde H^{(3)}$
lead to the 9 massive degrees of freedom given in Table \ref{table2}.
These massive fields do not appear in the potential, so we do not
consider them in this analysis.
The complete equation of motion for $\tilde B^{(3)}$ reads:
\begin{multline}
  \tfrac{1}{12}\nabla_{\rho}
  \Big(
    \E^{-\phi}\tilde H^{(4)}{}^{\mu\nu\lambda\rho\,\beta\gamma\delta}
    G_{\alpha_1\beta} G_{\alpha_2\gamma}G_{\alpha_3\delta}
  \Big)
  \\
  - \varepsilon^{\mu\nu\lambda\rho}
    {\bar D}_\rho A^I{}_{[\alpha_1} {\cal F}^I{}_{\alpha_2\alpha_3]}
  + \tilde H^{(3)}{}^{\mu\nu\lambda\,\beta\gamma}
    G_{\beta\delta}G_{\gamma[\alpha_1} f^{\delta}{}_{\alpha_2\alpha_3]} = 0 \,,
\label{eom-B3}
\end{multline}
while for $\tilde B^{(4)}$ we find
\begin{equation}
  \tfrac{1}{2}\varepsilon^{\mu\nu\lambda\rho}
    \mathcal{F}^I{}_{[\alpha_1\alpha_2}{\cal F}^I{}_{\alpha_3\alpha_4]}
  + \tilde H^{(4)}{}^{\mu\nu\lambda\rho}{}_{\beta[\alpha_1\alpha_2}
    f^{\beta}{}_{\alpha_3\alpha_4]} = 0\,,
\label{eom-B4}
\end{equation}
where the indices on $\tilde H^{(4)}$ have been lowered with 
$G_{\alpha\beta}$. Equation (\ref{eom-B4})
can be rewritten with the use of (\ref{redefH4}):
\begin{equation}
  \tfrac{1}{2}\mathcal{F}^I{}_{[\alpha_1\alpha_2}\mathcal{F}^I{}_{\alpha_3\alpha_4]}
  + \E^\phi\,\mathcal{H}_{\beta[\alpha_1\alpha_2}f^{\beta}{}_{\alpha_3\alpha_4]}
  =0\,.
\label{eom-B4r}
\end{equation}
Note that for  $\mathcal{G}=SU(2)\times SU(2)$
(\ref{eom-B4r}) does not constrain $\mathcal{H}_{abc}$ or
$\mathcal{H}_{\bar a\bar b\bar c}$. These are the fields for which we will
turn on a nonzero vev, following Section \ref{example}.

To proceed we 
make an additional simplification. We truncate
the scalar sector to only the singlets of $\mathcal{G}$. That implies 
\begin{subequations}
\begin{align}
  G_{ab}&= G_1\,\delta_{ab}\,,
  \quad G_{\bar a\bar b}=G_2\,\delta_{\bar a\bar b}\,,
  \quad G_{a\bar a} = 0\,,
  \label{Gsinglet}
  \\
  A^I{}_{\alpha}&=0\,.
\end{align}
\end{subequations}
This truncation is well known in the analysis of scalar potentials: an
extremum in the singlet sector is also an extremum of the potential without
truncation \cite{NPW}. The equations of motion (\ref{eom-B3})
now simplifies, and becomes
\begin{equation}
  \tfrac{1}{12}\nabla_{\rho}
  \big(
    \varepsilon^{\mu\nu\lambda\rho}
    \mathcal{H}_{\alpha_1\alpha_2\alpha_3}
  \big)
  + \tilde H^{(3)}{}^{\mu\nu\lambda}{}_{\delta[\alpha_1}
    f^{\delta}{}_{\alpha_2\alpha_3]} = 0\,.
\label{eom-B3r}
\end{equation}
For the choice $\alpha_1\alpha_2\alpha_3=abc$ the
second term in (\ref{eom-B3r}) vanishes and 
${\cal H}_{abc}$, and similarly ${\cal H}_{\bar a\bar b\bar c}$, is constant.
In this truncation we may therefore use the simplified analysis at the beginning
of this section, and are again led to the potential (\ref{Vflux}).

In the analysis of the potential we write
\begin{equation}
  \mathcal{H}_{abc}=\mathcal{H}_1\,\varepsilon_{abc}\,,\quad
  \mathcal{H}_{\bar a\bar b\bar c}=\mathcal{H}_2\,\varepsilon_{\bar a\bar b\bar c}\,.
\end{equation}
In terms of $G_i$ and ${\cal H}_i$, $i=1,2$ we find
\begin{equation}
  V_{{\rm singlet}} = \E^\phi 
  \Big( - \tfrac{3}{2}\big[ (G_1)^{-1}+(G_2)^{-1} \big]
        + \tfrac{1}{2}\big[ (G_1)^{-3}(\mathcal{H}_1)^2 
                             + (G_2)^{-3}(\mathcal{H}_2)^2 \big]
  \Big)\,.
\label{Vsinglet}
\end{equation}
The minimum value for fixed $e^\phi$ is at $G_1 = \mathcal{H}_1$, $G_2 = \mathcal{H}_2$ 
and reads
\begin{equation}
  V_0 = -\E^\phi\left(\frac{1}{|\mathcal{H}_1|}+\frac{1}{|\mathcal{H}_2|}\right).
\end{equation}
The potential in this extremum with respect to the $G_i$ 
is a negative function of the dilaton.

\section{Comparison with gauged $N=4$ supergravity and conclusions\label{conclusions}}

We performed the Scherk-Schwarz reduction of the 6-form 
version of $N=1,\ d=10$ supergravity. The result (\ref{finalL})
contains a scalar potential that receives contributions
of the form (\ref{Vflux}) from the vevs $\mathcal{H}_{\alpha\beta\gamma}$ 
of certain components of the 3-form fields $\tilde B^{(3)}$. 
In the reduction of the 2-form version the
potential takes on a similar form \cite{KM}. There the ${\cal H}$
are constant fluxes of the 2-form along cycles of the internal space.
We analyzed in particular the case $\mathcal{G}=SU(2)\times SU(2)$
and verified that, in the restriction to singlets of the gauge group, 
the potential (\ref{Vsinglet}) is identical to that of \cite{KM}.
This potential has a minimum in the directions parametrized 
by $G_1$ and $G_2$. 
In contrast with the torus reduction -- where
the 6-form yields 15 massless scalars and 12 massless vectors (see Appendix \ref{T6-B6}) --
the Scherk-Schwarz reduction over $SU(2)\times SU(2)$ yields only
massive degrees of freedom: six 2-forms and nine 3-forms. 
This is also the case in the reduction of the ten-dimensional 2-form
(see Appendix \ref{B2}).
The above is with the exception of the axion, which is massless in all these cases
and does not appear in the scalar potential.
The stabilization of the dilaton and the axion requires methods beyond the
Scherk-Schwarz reduction used in this paper.

We now want to compare our results, in the absence of Yang-Mills fields,
to those of gauged $N=4$ supergravity in
four dimensions. This comparison is facilitated by the analysis of
\cite{KM}, where the gauge groups associated with the Scherk-Schwarz
and flux reductions were identified. 
In the absence of Yang-Mills fields there are 12 generators, satisfying 
the algebra ($i,j=1,\ldots,6$):
\begin{equation}
\begin{split}
  [T_i,T_j] &= f_{ij}{}^k T_k +  \beta \,f_{ij}{}^k P_k\,,
  \\
  [T_i,P_j] &= f_{ij}{}^k P_k \,,
  \\
  [P_i,P_j] &= 0 \,.
\end{split}
\label{algebra}
\end{equation}
For $\beta=0$ this is the algebra associated with the Scherk-Schwarz reduction,
$\beta$ introduces the 3-form fluxes.
For $i,j=1,2,3$, and $f_{ij}{}^k=\varepsilon_{ijk}$ it corresponds to the group
$CSO(3,0,1)$ \cite{CSO}.

We start with the formulation discussed
in \cite{DWP,DWPT}, where we limit ourselves to the case of gauged 
supergravity with six additional vector multiplets. This field content
corresponds to that of the  reduction discussed in Section \ref{sugra}.
The scalars in \cite{DWP,DWPT} are the dilaton and axion in the
$SU(1,1)/U(1)$ coset, and
scalars which parametrize an $SO(6,6)/SO(6)\times SO(6)$ coset.
These are expressed in terms of $Z_a{}^{R}$, 
$R=1,\ldots,12,\ a=1,\ldots,6$, with constraint 
$Z_a{}^{R}\eta_{RS}Z_b{}^{S}=-\delta_{ab}$. Here $\eta_{RS}$ is the
$SO(6,6)$ invariant metric, chosen to be 
${\rm diag}\,(-1,\ldots,-1,1,\ldots,1)$. The scalar potential reads
\begin{equation}
  V_0 = \tfrac{1}{4}\E^\phi
        \Big(Z^{RU}Z^{SV} \big( \eta^{TW}+\tfrac{2}{3}Z^{TW} \big)\Big)
        f_{RST}f_{UVW}\,,
\label{potPW0}
\end{equation}
where $Z^{RS} = Z_a{}^{R}Z_a{}^{S}$ and $f_{RST}=\eta_{RU}f^U{}_{ST}$. 
Note that $Z^{RS}$ is not an element of $SO(6,6)$, but
$W^{RS} = \eta^{RS}+2 Z^{RS}$ is. 

To establish the precise correspondence between \cite{KM} and \cite{DWP,DWPT},
one has to identify the fields and structure constants.
In \cite{KM} the $SO(6,6)$ structure is described relative to the metric
\begin{equation}
  L =   \left(\begin{array}{cc}
           0 & \unitm{6} \\
         \unitm{6} & 0
       \end{array}\right)\,.
\end{equation}
The conversion to the metric $\eta$ involves $J$:
\begin{equation}
  J = \frac{1}{\sqrt{2}}
      \left(\begin{array}{cc}
          \unitm{6} & -\unitm{6} \\
         -\unitm{6} & -\unitm{6} 
      \end{array}\right)\,,\qquad \eta = J L J^{-1}\,.
\end{equation}
Similarly, the scalar matrix $M$ of \cite{KM} is related to $W=\eta+2Z$ by 
$W=JMJ^{-1}$. The identification is (see e.g. \cite{MS})
\begin{equation}
  Z_a{}^R = 
  \half\Big( E^{-1}(1+B)+E^T \;, \; -E^{-1}(1-B)+ E^T \Big) \,,
\end{equation}
where $E$ and $B$ are $6\times 6$ matrices. $G = EE^T$ are the scalars 
from the reduction of the metric and $B$ the scalars
from the 2-form. 
The structure constants in (\ref{algebra}) should also be
transformed by multiplication with $J$. For the gauge groups used in \cite{KM}
one always (also when including the Yang-Mills contribution) has the property
\begin{equation}
  \eta^{RU}\eta^{SV}\eta^{TW}f_{RST}f_{UVW} = 0\,.
\end{equation}
Thus the results of \cite{KM} all fit into the formulation of
\cite{MdRPW1}. 

We now choose the structure constants $f_{ij}{}^k$ in 
(\ref{algebra}) to be those of 
$SU(2)\times SU(2)$. 
The potential (\ref{potPW0}), 
restricted to singlets ($B=0$, $G$ as in (\ref{Gsinglet})) then
becomes
\begin{equation}
  V_1 = \E^\phi(V_{11}+V_{22})\,,
\label{PotPW1}
\end{equation}   
where
\begin{equation}
  V_{ii} = -\tfrac{1}{8}\big( 3(G_i)^{-1} - (\beta_i)^2(G_i)^{-3} \big)\,. 
\end{equation}
The parameters $\beta_i$ are the same as the flux parameter $\beta$ in (\ref{algebra}), 
we now have one of these for each factor of the gauge group. 
This is the same (up to a normalization) as the Scherk-Schwarz potential
including the vevs $\mathcal{H}_i$, if we identify 
$\beta_i={\cal H}_i$. We see therefore that this case
can be reproduced directly by a gauging in four dimensions.

However, these reductions do not give all gaugings in four dimensions. 
In the gauging of a direct product group $\mathcal{G}_{\rm gauge}$ in $N=4$,
$d=4$ supergravity, an $SU(1,1)$ angular parameter $\alpha$ can be introduced 
for each factor of $\mathcal{G}_{\rm gauge}$ \cite{MdRPW1}. 
These change in particular the dependence of the potential (\ref{potPW0})
on the dilaton and the axion. We considered this in detail for the gaugings
of semisimple groups in \cite{DWP,DWPT}. The Scherk-Schwarz reductions however
yield non-semisimple gauge groups and it is interesting to see the effect
of $SU(1,1)$ angles on such gaugings.
The reduction over $SU(2)\times SU(2)$ yields two copies of (\ref{algebra})
with $f_{ij}{}^k=\varepsilon_{ijk}$, i.e.
$\mathcal{G}_{\rm gauge}=CSO(3,0,1)\times CSO(3,0,1)$.
We can thus introduce two angles $\alpha_i$ and find the following 
potential ($\ell$ is the axion):
\begin{equation}
\begin{split}
  V_2 
  &= \E^\phi \Big( V_{11}(\ell\sin\alpha_1 - \cos\alpha_1)^2
                   +V_{22}(\ell\sin\alpha_2 - \cos\alpha_2)^2
  \\
  &\qquad\qquad\qquad\qquad\qquad
  + \E^{-2\phi}(V_{11}\sin^2\alpha_1 + V_{22}\sin^2\alpha_2) \Big)
  \\
  &\qquad\qquad\qquad\qquad\qquad
  - \frac{1}{4\sqrt{G_1G_2}}\sin(\alpha_1-\alpha_2)
    \bigg( 3- \frac{\beta_1}{G_1} \bigg)
    \bigg( 3- \frac{\beta_2}{G_2} \bigg)\,.
\label{PotPW2}
\end{split}
\end{equation}
For $\alpha_1=\alpha_2=0$ this reduces to (\ref{PotPW1}). For other values of the angles 
an extremum can be found for dilaton and axion if
\begin{equation}
  \Delta = 4\sin^2(\alpha_1-\alpha_2)\,V_{11}V_{22}>0\,.
\end{equation}
If this condition is satisfied the potential in the dilaton-axion extremum takes on the 
form
\begin{equation}
  V_{2,\ell,\phi} 
  = S \,\sqrt{\Delta}
    - \frac{1}{4\sqrt{G_1G_2}} \sin(\alpha_1-\alpha_2)
      \bigg( 3-{\beta_1\over G_1} \bigg)
      \bigg( 3-{\beta_2\over G_2} \bigg)\,,
\end{equation}
where $S$ is the sign of $V_{11}\sin^2\alpha_1+V_{22}\sin^2\alpha_2$. 
This potential has an extremum at 
$G_1=\beta_1$, $G_2=\beta_2$, the same values 
as for (\ref{Vsinglet}). 
The value of the potential in this extremum is:
\begin{equation}
  V_{{\rm min}} 
  = \frac{1}{\sqrt{\beta_1\beta_2}}
  \Big( \half S\, |\sin(\alpha_1-\alpha_2)| 
        - \sin(\alpha_1-\alpha_2) \Big)\,.
\end{equation}
This can be positive or negative for appropriate choices of the angles $\alpha_1$ and
$\alpha_2$. In the extremum the value of $S$ is negative. The extremum is at 
a positive value of $V$ for $\sin(\alpha_1-\alpha_2)<0$. In this case 
all second derivatives of the potential in the extremum
vanish. Expanding around the extremum one finds that in the $G_1$ and $G_2$ directions
the leading term  is cubic function of $G_i-2$, while in mixed direction the potential
is quartic and has a maximum. We emphasize the point that 
in contrast to (\ref{Vsinglet}) the potential (\ref{PotPW2})
has an extremum in all variables.

In this paper we have restricted the analysis of the potentials to the singlet sector.
This truncation has the property that extrema in these singlet variables are
also extrema in the full theory. Nevertheless, the singlets do not determine the
full structure, and it would be interesting to extend the analysis to all fields.

We have found that the results of the reduction of the 
$N=1$ $d=10$ theories in ten dimensions of \cite{KM} and of this paper are 
related to a gauged $N=4$ supergravity in four dimensions. In particular, 
the scalar potentials are the same. This does not necessarily imply that the four
dimensional theories are equivalent. From the reduction we find massive two- and 
three-forms. Although the number of degrees of freedom is the same as in gauged
$N=4$ supergravity, the degrees of freedom are represented by different fields.
It would be interesting to develop an $N=4$ supergravity in terms of higher-form
fields directly in four dimensions and to consider the possible gaugings. 

The gauge group (\ref{algebra}) is not semisimple, and therefore the analysis 
of \cite{DWP,DWPT}
does not apply. It will be interesting to systematically extend this analysis to
more general groups. It remains an interesting problem to establish a relation
between string theory and the $SU(1,1)$ angles in $d=4$.

\acknowledgments

\bigskip

The work of MGCE and DBW is part of the research program of the
``Stichting voor Fundamenteel Onderzoek van de Materie'' (FOM).
This work is supported in part by the European Commission RTN
program HPRN-CT-2000-00131, in which the Centre for Theoretical
Physics in Groningen is associated to the University of Utrecht.
SP thanks the University of Groningen for hospitality. DBW thanks
Peter~W.~Michor for kindly answering questions about Lie group cohomology.

\appendix

\section{Notation and conventions}\label{conventions}

Our line element is 
$\dd s^2=g_{\mu\nu}\dd x^\mu \otimes \dd x^\nu=\eta_{ab}e^a\otimes e^b$
with $\eta_{ab}={\rm diag}(-1,+1,+1,\ldots)$. The $e^a$ are
vielbein 1-forms, $e^a=e^a_\mu \dd x^\mu$. The permutation symbol $\tilde\ve$ 
is defined by
\begin{equation}
  \tilde\ve_{12\ldots D}=+1,\quad \tilde\ve^{12\ldots D}=-1 \ .
  \label{eq-1}
\end{equation}
The Levi-Civit\`a tensor is defined by:
\begin{equation}
  \ve_{\mu_1 \ldots \mu_D}= e\,\tilde\ve_{\mu_1 \ldots \mu_D} \,,\quad
  \ve^{\mu_1 \ldots \mu_D}= e^{-1}\, \tilde\ve^{\mu_1 \ldots \mu_D}\,,
  \quad{\rm with}\ e^2=-\det{g}\,.
\end{equation}
$p$-forms $A^{(p)}$ are normalized as follows:
\begin{equation}
  A^{(p)} 
  = \tfrac{1}{p!}A_{\mu_1\ldots\mu_p}\,\dd x^{\mu_1}\w\ldots\w \dd x^{\mu_p}
  = \tfrac{1}{p!}A_{a_1\ldots a_p}\,e^{a_1}\w\ldots\w e^{a_p}\,,
\end{equation}
where 
\begin{equation}
  \dd x^{\mu_1}\w\ldots\w \dd x^{\mu_p}
  = \dd x^{\mu_1}\otimes\ldots\otimes\dd x^{\mu_p}\pm\mbox{permutations}.
\end{equation}
The exterior derivative $\dd$ acts as:
\begin{equation}
  (\dd A)_{\mu_1\ldots\mu_{p+1}}
  =(p+1)\partial_{[\mu_1}A_{\mu_2\ldots\mu_{p+1}]}\,.
\end{equation}
The Hodge dual ${\star A}^{(p)}$ is defined by:
\begin{equation}
  ({\star A})_{\mu_1\ldots\mu_q}
  = \tfrac{1}{p!}\ve_{\mu_1\ldots\mu_q}{}^{\nu_1\ldots\nu_p}A_{\nu_1\ldots\nu_p}\,,
\end{equation}
where $q=D-p$. Thus $(\star)^2=(-)^{pq-1}$ and 
\begin{equation}
  {\star A}^{(p)}\w B^{(p)}
  = \tfrac{1}{p!}A_{\mu_1\ldots\mu_p}B^{\mu_1\ldots\mu_p}\,{\star\mathbf{1}}\,,
\end{equation}
where ${\star\mathbf{1}}=e\,\dd x^1\w\ldots\w \dd x^D$ is the volume form.
We follow the conventions of \cite{KM} for the Riemann tensor.

\section{Examples of the cohomology\label{furthercoho}}

\subsection{6-Torus reduction of the 6-form\label{T6-B6}}

The 6-torus is a group manifold, $T^6=U(1)^6$, 
hence the analysis of Section \ref{coho} can be applied. 
The cohomology of $T^6$ is generated by the left-invariant
1-forms $\dd y^i\in\mathfrak{u}(1)^*$, $i=1,\ldots,6$. We have
\begin{eqnarray}
  && b_p(T^6)=\left(\begin{array}{c}6\\p\\ \end{array}\right)\,,
  \nonumber\\
  && {\rm dim}\,C^{(n)} = 0\,,\quad n=0,\ldots,6\,.
\end{eqnarray}
There are no local shift symmetries, therefore only massless states will occur.
There are 15 (scalar) states from $B^{(2)}$, 12 states from $B^{(1)}$ and
one axion from $B^{(0)}$, as can be read off from Table
\ref{table1}. The massless 2-forms can be dualized to scalars
in four dimensions.

The number of possible vevs for $B^{(3)}$ is 20, which are
precisely those in the span of $(\bigwedge^3 \dd y^i)$.

\subsection{$\mathcal{G}=SU(2)\times U(1)^3 $ reduction of the 6-form\label{T3-B6}}

\begin{table}
\begin{center}
\label{table3}
  \begin{tabular}[tbp]{|c|l|c|c|c|}
		\hline field & features & dimension & \# DOF & total DOF\\
		\hline
		\ $B^{(4)}$ &  gauge  & $3$ &  0  & 0  \\
		\ &  massive          & $9$ &  0  & 0  \\
		\ &  massless         & $3$ &  0  & 0  \\
		\hline
		\ $ B^{(3)}$ &  gauge & $9$ &  0  & 0  \\
		\ &  massive          & $9$ &  1  & 9  \\
		\ & massless          & $2$ &  0  & 0  \\
		\hline
		\ $B^{(2)}$ &  gauge  & $9$ &  0  & 0  \\
		\ &  massive          & $3$ &  3  & 9  \\
		\ &  massless         & $3$ &  1  & 3  \\
		\hline
		\ $B^{(1)}$ &  gauge  & $3$ &  0  & 0 \\
		\ &  massive          & $0$ &  3  & 0  \\
		\ &  massless         & $3$ &  2  & 6  \\
		\hline
		\ $B^{(0)}$ & gauge   & $0$ &  0  & 0  \\
		\ & massive           & $0$ &  0  & 0  \\
		\ & massless          & $1$ &  1  & 1  \\
		\hline
	\end{tabular}
\caption{{\small
  Result of the analysis of the
  Scherk-Schwarz gauge transformations of the Kalb-Ramond fields
  $B^{(n)}$ for the group $SU(2)\times U(1)^3$, for the
  reduction of $\hat B^{(6)}$ from 10 to 4 dimensions.
  See the caption
  of Table \ref{table1} and \ref{table2} for more details.
}}
\end{center}
\end{table}

The Betti numbers of $SU(2)$ and $U(1)^3$ are given by
\begin{equation}
  (b^{SU(2)}_0,\ldots,b^{SU(2)}_3) = (1,0,0,1),\quad
  (b^{U(1)^3}_0,\ldots,b^{U(1)^3}_3) = (1,3,3,1)
\end{equation}
and we use Kunneth's formula to obtain the Betti numbers of
$\mathcal{G}=SU(2)\times U(1)^3$:
\begin{equation}
  b^{\mathcal{G}}_{r}=\sum_{n+m=r}b^{SU(2)}_m b^{U(1)^3}_n,\quad
  (b^{\mathcal{G}}_{0},\ldots,b^{\mathcal{G}}_{6}) = (1,3,3,2,3,3,1).
\end{equation}
The resulting degrees of freedom are given in Table \ref{table3}. Of course
the total number of degrees of freedom, the sum of the numbers
in the last column in Table \ref{table3}, is again 28.

There are 9 massive degrees of freedom in $B^{(3)}$.
The two massless components in $B^{(3)}$ allow us to introduce
vevs, they correspond to the 3-forms
\begin{equation}
  \sigma^1\sigma^2\sigma^3, \, \quad \mbox{ and } \quad
  \dd y^1 \dd y^2 \dd y^3\,,
\end{equation}
where $\sigma^a\in\mathfrak{su}(2)^*$ and $\dd y^i\in\mathfrak{u}(1)^*$.
The massless components of  $B^{(4)}$ do not appear in the action and do not
give rise to vevs.
There are 3 massive 2-forms, as well as 3 massless ones.
The latter can be dualized to scalars and may contribute to the potential.
There are three vectors that can be gauged away, corresponding to
$SU(2)$, and three massless physical vectors corresponding to $U(1)^3$.
The single scalar $B^{(0)}$ is the axion.

\subsection{$\mathcal{G}=SU(2)\times SU(2)$ reduction of the 2-form\label{B2}}

\begin{table}
\begin{center}
\label{table4}
  \begin{tabular}[tbp]{|c|l|c|c|c|}
		\hline field & features & dimension & \# DOF & total DOF\\
		\hline
		\ $B^{(2)}$ &  gauge  & 0   &  0  & 0  \\
		\ &  massive          & 0   &  0  & 0  \\
		\ &  massless         & 1   &  1  & 1  \\
		\hline
		\ $ B^{(1)}$ &  gauge & 0   &  0  & 0  \\
		\ &  massive          & 6   &  3  & 18 \\
		\ & massless          & 0   &  2  & 0  \\
		\hline
		\ $B^{(0)}$ &  gauge  & 6   &  0  & 0  \\
		\ &  massive          & 9   &  1  & 9  \\
		\ &  massless         & 0   &  1  & 0  \\
		\hline
	\end{tabular}
\caption{{\small
  Result of the analysis of the Scherk-Schwarz gauge
  transformations of the Kalb-Ramond fields
  $B^{(n)}$ for the group $SU(2)\times SU(2)$, for the
  reduction of $\hat B^{(2)}$ from 10 to 4 dimensions.
  See the caption
  of Table \ref{table1} and \ref{table2} for more details.
}}
\end{center}
\end{table}

Another interesting example is the reduction of a 2-form field $\hat B^{(2)}$
from 10 to 4 dimensions \cite{KM,MS}. 
In this case we obtain 2-, 1-, and 0-forms:
$B^{(2)}$, $B^{(1)}{}_{\alpha}$ and $B^{(0)}{}_{\alpha\beta}$ respectively. 
We consider $\mathcal{G}=SU(2)\times SU(2)$, of which the
relevant Betti numbers are
\begin{equation}
  (b_0,\ldots,b_3) = (1,0,0,2)\,.
\end{equation}
The dimensions of the spaces $C^{(p)}$ are ${\rm dim}\,C^{(0)}={\rm dim}\,C^{(1)}=0$,
${\rm dim}\,C^{(2)}=6$, ${\rm dim}\,C^{(3)}=9$.
The resulting degrees of freedom are given in Table \ref{table4}. 
There is a massless 2-form (which can be dualized to an axionic scalar), 
there are six massive vectors, and nine
massive scalars, giving again a total of 28 degrees of freedom.

These are the same number of degrees of freedom as for the 6-form 
case (Table \ref{table2}): the massless now appears as
a two-form, the 9 degrees of freedom represented by 3-forms in Table
\ref{table2} now appear as scalars, the massive 2-forms
from Table \ref{table2} now appear as vectors.

For the $T^6$ reduction of the 2-form version we have
$(b_0,b_1,b_2)=(1,6,15)$, giving massless degrees of
freedom only: 1 for $B^{(2)}$ (the axion), 6 massless
vectors for $B^{(1)}$, and 15 massless scalars for
$B^{(0)}$. Again these are the same degrees of freedom
as obtained from a reduction of the 6-form
(see Appendix \ref{T6-B6}).

\end{document}